\documentclass[11pt,twoside]{article}

\usepackage{asp2006} \usepackage{epsf} \usepackage{lscape}
\usepackage{graphicx}

\begin{document}
\title{Extragalactic Classical Nova Surveys}

\author{Matthew~J.~Darnley$^1$,~Diego~T.~R.~Black$^1$, Michael~F.~Bode$^1$,
  Andrew~M.~Newsam$^1$,
  Eamonn~Kerins$^2$, Tim~J.~O'Brien$^2$,
  Karl~Misselt$^3$, and Allen~W.~Shafter$^4$}
 \affil{$^1$Astrophysics Research Institute, Liverpool
  John Moores University, UK\\
$^2$Jodrell Bank Observatory, University of Manchester, UK\\
$^3$University of Arizona, USA\\
$^4$San Diego State University, USA}


\begin{abstract}
We are currently involved in a multifaceted campaign to study
extragalactic classical novae in the Local Group and beyond.  Here we
report on-going results from the exploitation of the POINT-AGAPE M31
dataset; initial results from our Local Group imaging, and
spectroscopic CNe follow-up campaign and introduce the Liverpool
Extragalactic Nova Survey.
\end{abstract}

\section{Introduction}

Although much has been learnt from the study of Galactic classical
novae (CNe), it is clear that Galactic data are not ideal for
establishing the population characteristics of novae because these
data are often heavily biased by selection effects.  To-date no RNe
have been identified outside the Milky Way and its companions, but CNe
have been studied in about a dozen galaxies.  To gain further insight
into the population of novae, and specifically to explore further the
question of whether there exist two distinct nova populations
\citep[see e.g.][]{1992A&A...266..232D}, we are involved in a number
of campaigns to study a large sample of novae in the Local Group and
beyond.  The following sections briefly describe the current status of
the three parts of our extragalactic CN work: the POINT-AGAPE survey;
our Local Group CN follow-up project, and the Liverpool Extragalactic
Nova Survey.

\section{The POINT-AGAPE Survey}

The Pixel-lensing Observations with the Isaac Newton Telescope --
Andromeda Galaxy Amplified Pixels Experiment (POINT-AGAPE) survey
(Calchi Novati et al. 2003) was an optical search for gravitational
microlensing events towards the Andromeda Galaxy (M31).  As well as
microlensing, the survey was sensitive to many different classes of
variable stars and other transients, including Classical Novae
\citep{2002AIPC..637..481D}.

Previous work with the POINT-AGAPE dataset included the development of
an automated CN detection pipeline, which led to the discovery of 20
CNe \citep{2004MNRAS.353..571D}.  Using the results from the
catalogue, a global CN rate for M31 of $65^{+16}_{-15}$~yr$^{-1}$ was
derived \citep{2006MNRAS.369..257D}.  Separate M31 bulge and disk
rates of $38^{+15}_{-12}$~yr$^{-1}$ and $27^{+19}_{-15}$~yr$^{-1}$
respectively were also determined.  The derived global rate is a
factor of around two higher than the most robust previous result of
$37^{+12}_{-8}$ \citep{2001ApJ...563..749S} and strong evidence in
favour of two separate CN populations was provided: one associated
with the M31 bulge, the other with the disk.

Darnley et al. (2006) were able to use the M31 dataset and
various assumptions about the Milky Way
\citep[see][]{2002AIPC..637..462S} to deduce a Galactic bulge nova
rate of $14^{+6}_{-5}$~yr$^{-1}$, a disk rate of $20^{+14}_{-11}$
yr$^{-1}$ and a global Galactic rate of $35^{+15}_{-12}$~yr$^{-1}$.
This rate is remarkably similar to independent estimates computed from
direct observations of the Milky Way's CN population
\citep{1997ApJ...487..226S}.

Recently, an additional fourth season of POINT-AGAPE legacy data has
been obtained.  These data are being analysed with the expectation of
additional CNe detections and hence further refining the M31 bulge and
disk rates, and strengthening the evidence in favour of two distinct
M31 CN populations.

\section{The Local Group}

As part of ongoing work observing Local Group CNe, programmes with
guaranteed observing time are in place on a number of telescopes to
follow-up novae discovered in the Andromeda Galaxy, its companion
(M32) and the Triangulum Galaxy (M33).  To provide optical follow-up
observations the 1m telescope at the Mount Laguna Observatory (MLO),
the Steward 2.3m and the LT are employed.  Time has also been granted
on the Hobby-Eberly Telescope (HET) to obtain low-resolution optical
spectra and Spitzer Space Telescope time to perform IR photometry and
spectroscopy.

\begin{figure}
\center\includegraphics[keepaspectratio=true,clip=true,angle=270,width=4in]{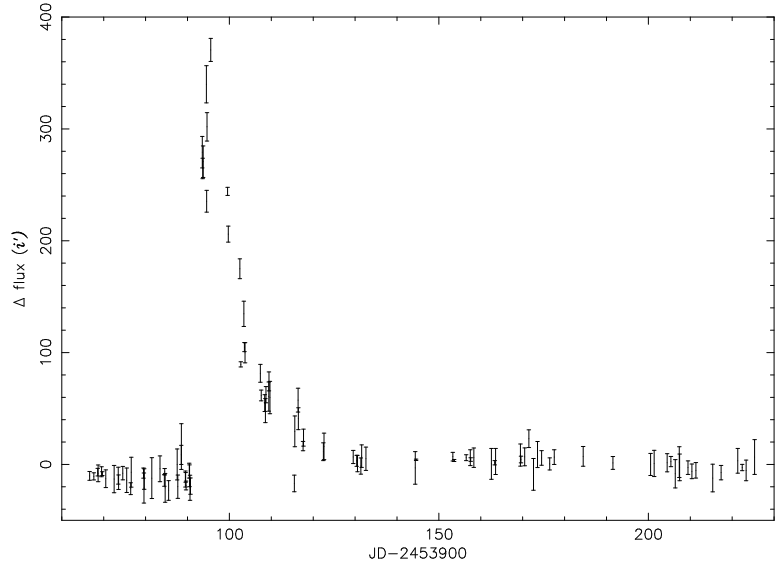}
\caption{Angstrom DIA light-curve of Classical Nova 2006 \#8
  \citep{2007ApJ...661L..45D}, first announced by
  Burwitz et al. (2006).  This nova has a $t_{2}\leq10$ days and
  is classed as a ``very fast'' nova.  This light curve provides the
  best sampled light-curve of an extragalactic CN to-date.}
\label{Angstrom_lightcurve}
\end{figure}

Systematic studies of M31 (and the Local Group) have become feasible
for the first time in recent years due in part to the advent of
robotic telescopes, such as the Liverpool Telescope
\citep[LT,][]{2004SPIE.5489..679S} and Faulkes Telescopes.  However a
large number of novae are discovered in M31 each year by amateurs and
professionals alike.  Over the past two years, a total of 31 CNe have
been discovered during the $\sim8$ month M31 observing season.  This
number neglects any contribution from routine optical imaging of M31
with the LT (undertaken by the Angstrom project,
\citep[see][and Figure~\ref{Angstrom_lightcurve}]{2006MNRAS.365.1099K}, MLO
and Steward, amongst others.  The Local Group nova sample is also
being supplemented by CN alerts from the Angstrom M31 bulge
micro-lensing survey \citep{2007ApJ...661L..45D,ATel1192} and
serendipitous nova discoveries made by the Lick Observatory Supernova
Search (LOSS) and the ROTSE IIIb programme.

\begin{figure}
\center\includegraphics[keepaspectratio=true,clip=true,width=4in,viewport=60
  15 430 285]{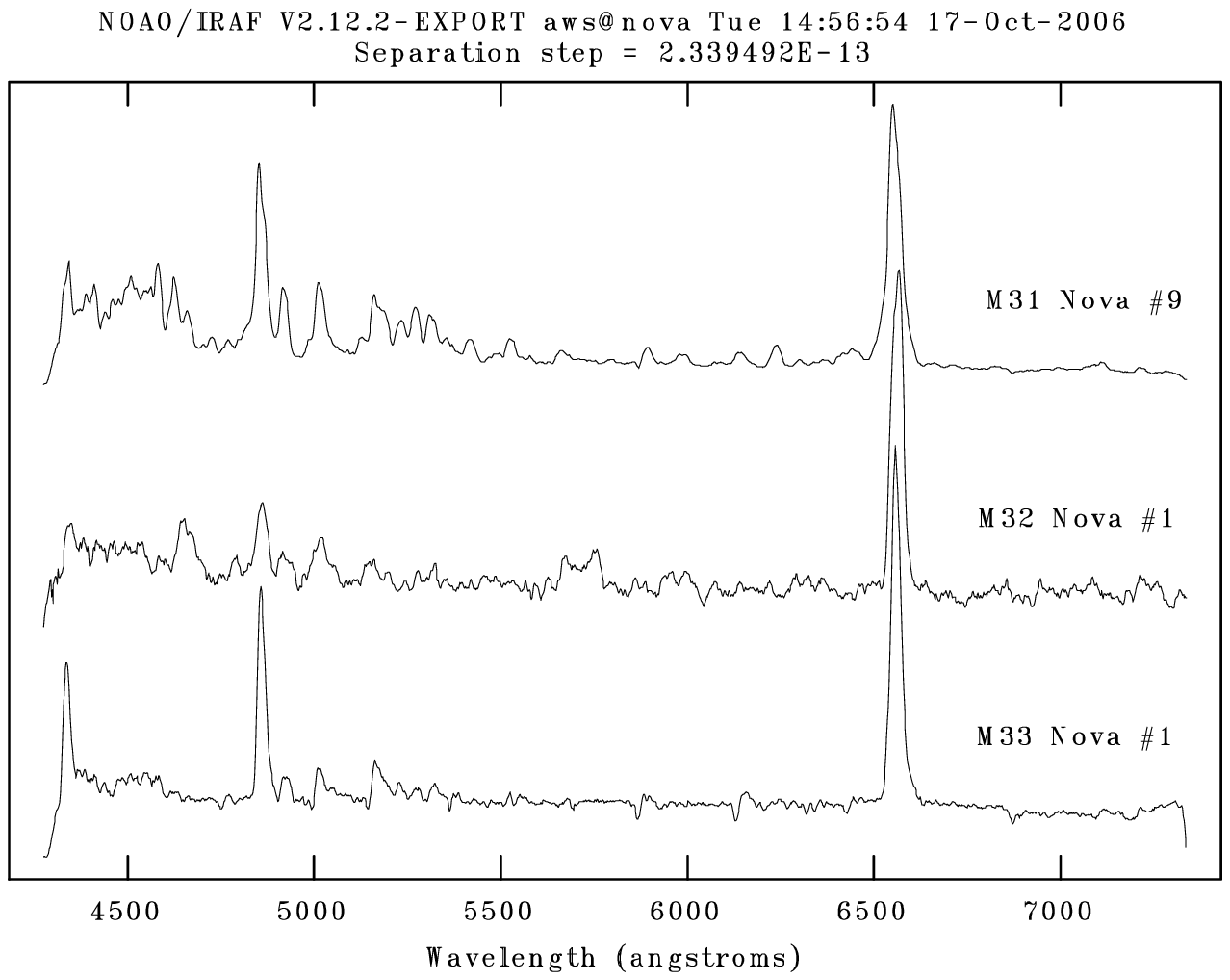}
\caption{HET spectra of three of our target Local Group novae, one
  from each of M31, M32 and M33 \citep{2006AAS...209.0920S}.  These
  spectra show strong Hydrogen emission lines as well as prominent Fe
  II emission lines with characteristic P Cygni profiles
  \citep{1992AJ....104..725W}.  All three are classic examples of Fe
  II CNe.}
\label{HET_spectra}
\end{figure}

During the last M31/M33 observing season (August 2006 -- February
2007) six novae in M31 and one in each of M32 and M33 have been
followed-up with all three optical telescopes and low-resolution
spectroscopy has been performed with the HET
\citep{2006ATel..923....1S}.  Figure~\ref{HET_spectra} shows HET
spectra of three Local Group novae.  Infrared photometry and
spectroscopy for four CNe in M31 has recently been obtained from
Spitzer; these data are still being analysed.

\section{Liverpool Extragalactic Nova Survey}

The Liverpool Extragalactic Nova Survey (LENS) is a high cadence
extragalactic CN monitoring survey.  Conceived to expand upon the
results of the POINT-AGAPE CN survey, LENS studies three more distant
galaxies, covering a range a Hubble types; namely, M81, NGC 2403 and
M64.  This survey operates primarily on the robotic 2m LT and also
utilises some RoboNet-1.0 time on the Faulkes Telescope North
\citep[FTN,][]{2007P&SS...55..582B}.  A primary objective is to
determine how the nova rate varies with Hubble type.

LENS has to-date taken three seasons of data (including an initial
pilot season during the commissioning of the LT) for each galaxy, with
guaranteed time on the LT to conduct a fourth observing season.  These
data are reduced using a fully automated difference-image-analysis
(DIA) pipeline \citep{Angstrom_Pipeline}, with nova detection via the
POINT-AGAPE algorithm \citep{2004MNRAS.353..571D}.  Variable object
detection and classification is currently being performed on the LENS
dataset, and a number of CN candidates have already been identified.

\acknowledgements

The Liverpool Telescope is operated on the island of La Palma by
Liverpool John Moores University in the Spanish Observatorio del Roque
de los Muchachos of the Instituto de Astrofisica de Canarias with
financial support from the UK Science and Technology Facilities
Council (STFC).  FTN is operated by the Las Cumbres Observatory Global
Telescope network.  AWS acknowledges support through NSF grant
AST-0607682.

\end{document}